\documentclass[final,3p,times,twocolumn]{elsarticle}
\usepackage{tabularx}
\usepackage{graphics}
\usepackage{graphicx}
\usepackage{amssymb}
\usepackage{amsthm}
\usepackage{amsmath}
\usepackage{upgreek}
\usepackage{subfigure}
\usepackage{wasysym}
\usepackage{textcomp}
\usepackage{array}
\usepackage{booktabs}
\usepackage{paralist}
\usepackage{verbatim}
\usepackage{float}

\biboptions{square}
\journal {Nuclear Instruments and Methods in Physics Research Section A}

\begin{document}
\begin{frontmatter}

\title{Cosmic test of sTGC detector prototype made in China for ATLAS experiment upgrade}

\cortext[cauthor]{Correspondence to: 72 Binhai Road, Jimo, Qingdao 2266237, Shandong, China.\\
E-mail address: duyanyan199034@163.com;duanyanyun1991@163.com}

\author[1:sdu]{Xiao Zhao}  
\author[1:sdu]{Dengfeng Zhang} 
\author[1:sdu]{Wenlong Li} 
\author[1:sdu]{Changyu Li} 
\author[1:sdu]{Chengguang Zhu}  
\author[1:sdu]{Han Li}
\author[2:ustc]{Shengquan Liu}   
\author[2:ustc]{Peng Miao}   
\author[3:ytu]{Yanyan Du\corref{cauthor}} 
\author[1:sdu]{Yanyun Duan\corref{cauthor}} 

\address[1:sdu]{Center for Particle Science and Technology, Institute of Frontier and Interdisciplinary Science, Shandong University, Qingdao 2266237, China}
\address[2:ustc]{Department of Modern Physics, University of Science and Technology of China, Hefei 230026, China}
\address[3:ytu]{Department of physics, Yantai University, Yantai 264005, China}

\begin{abstract}
 Following the Higgs particle discovery, the Large Hadron Collider complex will be upgraded in several phases allowing the luminosity to increase to 7$\times$10$^{34}$cm$^{-2}$s$^{-1}$. In order to adapt the ATLAS detector to the higher luminosity environment after the upgrade, part of the ATLAS muon end-cap system, the Small Wheel, will be replaced by the New Small Wheel. The New Small Wheel includes two kinds of detectors: small-strip Thin Gap Chambers and Micromegas. Shandong University, part of the ATLAS collaboration, participates in the construction of the ATLAS New Small Wheel by developing, producing and testing the performance of part of the small-strip Thin Gap Chambers. This paper describes the construction and cosmic-ray testing of small-strip Thin Gap Chambers in Shandong University.
\end{abstract}


\end{frontmatter}

\section{Introduction}

After the discovery of the Higgs particle\cite{ref:det1}, the LHC will be upgraded to the High Luminosity LHC(HL-LHC). To adapt the ATLAS detector to the high radiation environment, the ATLAS detector will be upgraded. The end-cap muon trigger system, located in the middle of the muon end-cap region, consists of the seven-layer TGC detectors. 90$\%$ of the Level 1 muon trigger cases produced by end-cap muon trigger system are fake triggers caused by the secondary particles created in the calorimeter and toroid magnetic system\cite{ref:det2}, which occupy more than half of the allowed bandwidth of the total level-1 trigger. For proper triggering on interesting physics events, it is necessary to reduce the fake trigger rate as low as possible\cite{ref:det2}. The ATLAS phase-1 upgrade will use the New Small Wheel (NSW) to substitute for the Small Wheel (SW) before the end-cap toroid magnetic system to combine with the exiting muon trigger Big Wheel to reduce the fake trigger rate. NSW includes two kinds of detectors: sTGC(small-strip Thin Gap Chamber) and MM(Micromegas). The NSW reconstructed muon segments with a precision of pointing direction of 1mrad are used to select muon tracks which point to the interaction point at Level 1 trigger, to suppress the fake trigger from the secondary particles. This requires a position resolution of 300$\mu$m in each sTGC quadruple station\cite{ref:det2}. The sTGC quadruplets are constructed in several sites by the NSW collaboration including Shandong University of China who will build quadruplet type S2. Before the mass production, a full-size porotype was constructed and tested with a cosmic test bed. Construction and performance measurement of the prototype is presented in this paper.

\section{ Construction of sTGC type S2 }

\begin{figure}
\begin{center}
\includegraphics[width=0.35\textwidth]{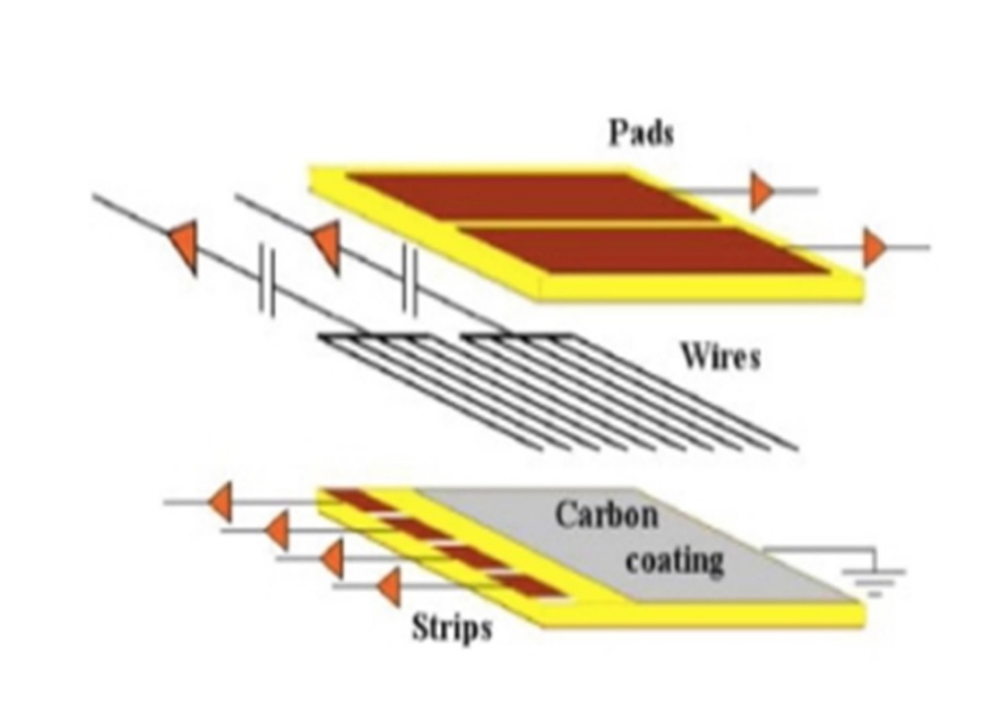}
\caption{The layout of sTGC type S2.}
\label{fig:sTGClayout}
\end{center}
\end{figure}

\begin{figure}
	\begin{center}
		\includegraphics[width=0.35\textwidth]{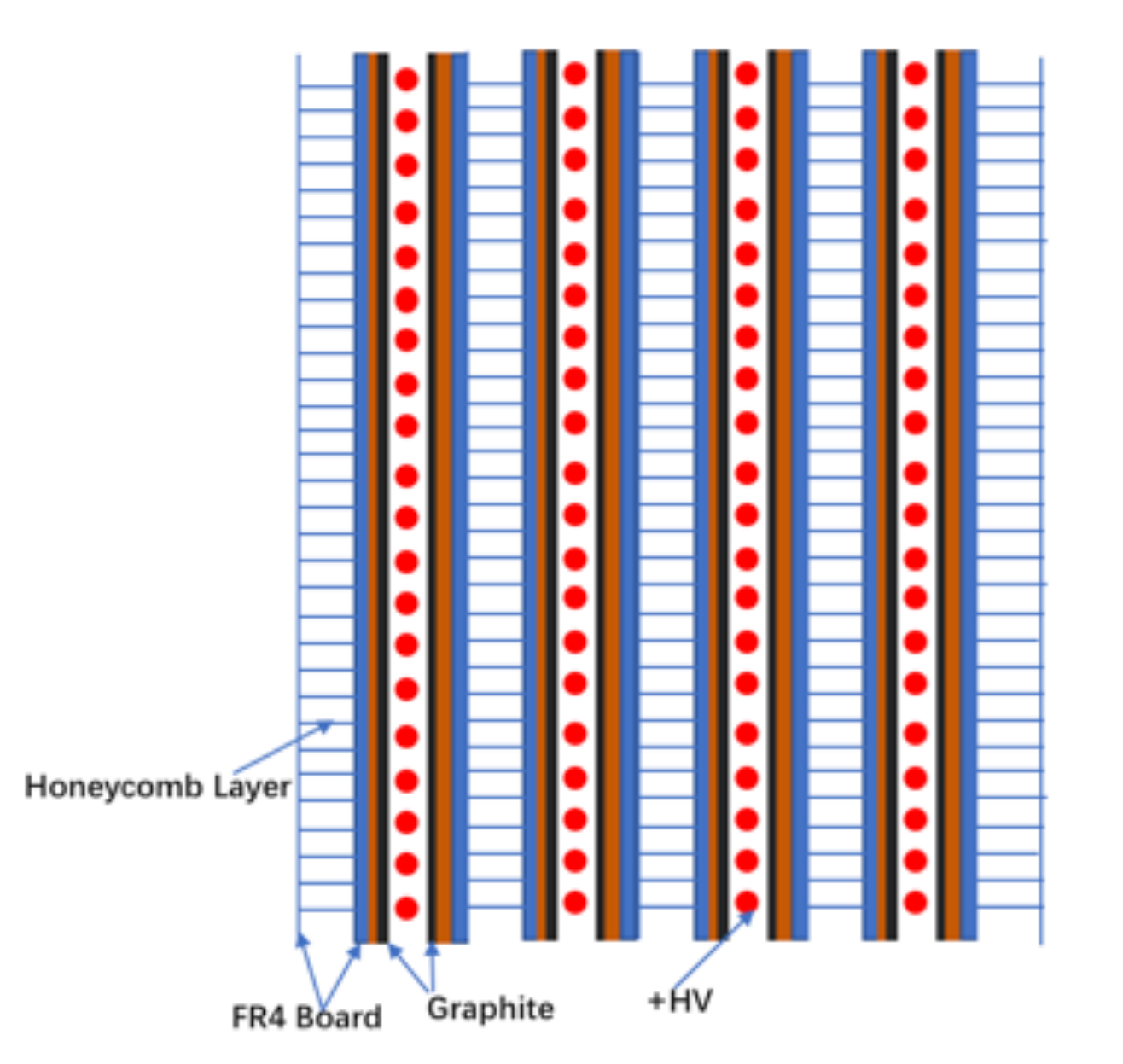}
		\caption{The layout of sTGC multiplet.}
		\label{fig:multipletlayout}
	\end{center}
\end{figure}

The schematic structure of the sTGC is shown in Figure~\ref{fig:sTGClayout}. The anode wire is made of gold-plated tungsten wire with diameter of 50 micrometers and spacing of 1.8mm, operating at 2900v. The anode wire is in the middle of the two cathode plates, and distances to the two cathode plates are both 1.4mm. The cathode is made of a 200$\mu$m thick FR4 insulation layer with graphite (resistance of around 200k$\Omega$/$\Box$ with a uniformity of +/-20$\%$) coating. Signal pick-up copper strips or pads are located behind the insulation layer and 1.3mm FR4 boards are the strong back of the thin insulation layer. A full size sTGC prototype of type S2 is trapezoidal，with the long base of 1087mm, short base of 731mm and height of 1191mm. Four chambers are glued together with five honeycomb layers sandwiched to formulate a rigid installation module, or quadruplet, as seen in Figure~\ref{fig:multipletlayout}.

 The graphite was sprayed on the cathode board using spraying machine and polished to around 200k$\Omega$ per square with a uniformity of +/-20$\%$.  The frames and inner supports are then glued onto the board to build a structure of so called half-chambers. On one of the half-chamber wires with a tension of 350 grams are wound and soldered on the frame. The half-chamber with wire is glued together with a half-chamber without wires to compose a sTGC chamber. The thickness of the periphery area of the chamber is controlled to be within a precision of 100$\mu$m by controlling the thickness of the board and frames. The planarity of the chamber is controlled to have a deviation below 100$\mu$m from a straight ruler attached on to the chamber surface. The chamber is tested for gas tightness and stability under a high voltage of 3200v between the anode wires and graphite layer and then scanned with x-ray radiation to find hot spot and the non-uniformity of the avalanche gain. Two chambers passing the above quality checks are glued into a doublet with a honeycomb layer sandwiched in between. The front-end board adaptors are then soldered on to the chamber connecting all the strip, pad and wire signals readout. The connectivity of the signal path is checked by reading signals from each channel when injecting a square wave into the wires. Two doublets are glued into a quadruplet with one honeycomb layer sandwiched in between. Two layers of honeycomb covered with a layer of cooper skin are glue to the outer side of the quadruplet for the purpose of protection and part of the Faraday cage. The front end board specific designed for sTGC are installed on to the quadruplet and the quadruplet is then inserted into the cosmic test bed\cite{ref:det3} for the performance test. 

\section{sTGC prototype performance test }

\subsection{Setup of sTGC test}

The cosmic test bed measures the arrival time and the hit position using scintillators and wire chambers\cite{ref:det3}. Four hits measured along the muon trajectory are used to reconstruct the track of a muon. The hit time and position with precision of $\sim$1cm and 0.7ns at the tested sTGC quadruplet are then calculated using the reconstructed track and on the fly time subtraction.

The VMM chip is developed by the NSW collaboration for the front-end signal processing of NSW detectors\cite{ref:det4}. Three prototypes of the front-end board (FEB) based on the VMM version 3 are installed onto the sTGC prototype. Due to a lack of the wire signal readout and a lack of enough front-end boards, signals from 3 layers of strips of the quadruplet are read out and the second coordinate of hit position is provided by the test bed. The measurements presented in this paper include: the noise and pedestal measurement of the integrated detector and FEB, the characteristic of the cluster of strip signals, the position resolution of the sTGC prototype, the relative rotation and shift of one of chamber taking the other two as reference, detection efficiency of the prototype.

\subsection{Noise and pedestal measurement}

\begin{figure}
	\begin{center}
\includegraphics[width=0.45\textwidth]{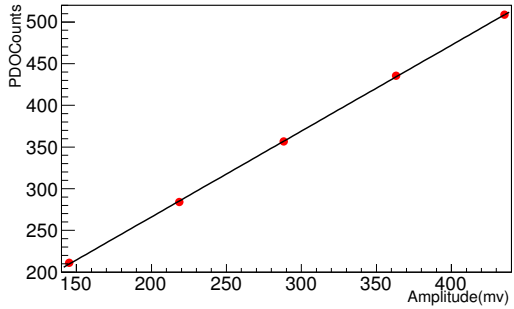}
		\caption{Relation of the measured analog amplitude and the PDO count.}
		\label{fig:PDOcount}
	\end{center}
\end{figure}

The charge in the signal is integrated, amplified and shaped, which can be monitored by oscilloscope from the monitoring pin of the chip or digitized to be read out as so called PDO count for offline analysis. As both methods are used and the result are cross checked in part of the test, the relation between the amplitude of the shaped signal and the PDO count are measured. Using the self-test functionality of the VMM chip, charge is injected into the VMM input and amplitude of the shaped signal is measured using both the oscilloscope and the digits of the PDO count. Adjusting the amount of the injected charge and comparing the measured analog amplitude and the PDO count, a proportional relation of them is expected. Fitting Figure~\ref{fig:PDOcount} with a straight line shows the amplitude of the shaped signal of 1mv corresponds to 1.03 PDO counts.

\begin{figure}
	\begin{center}
		\includegraphics[width=0.45\textwidth]{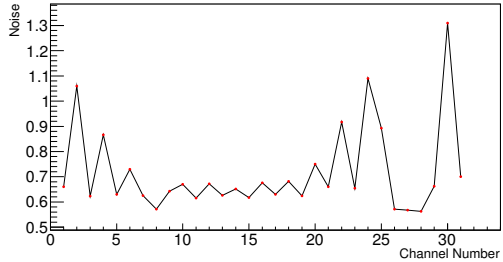}
		\caption{The noise of FEB with Adapter board.}
		\label{fig:noise}
	\end{center}
\end{figure}

Before measuring the noise of the integrated system of detector, adaptors and FEB, the noise of the FEB with only the adaptors connected is measured. Setting the gain to 1mv/fC and peaking time to 50ns, the analog signals from each channel are observed by the oscilloscope and the RMS of the magnitude of the analog signals are taken as the noise as shown in Figure~\ref{fig:noise}.

There are more than one thousand strip channels in the 3 chambers equipped with FEBs and it is time consuming to measure the noise one by one with an oscilloscope. Due to the fact that VMM chip only returns the signals which are over threshold and the threshold cannot be set to be 0, due to DAQ limitations, the strategy is to use the functionality of neighbor trigger of the VMM chip. The neighbor trigger functionality means signals below threshold will still be readout if the neighboring channel is over the readout threshold. The thresholds of all channels are set to be as low as possible and the charge is injected to one out of three channels. The other two channels which are neighbors of the charge injected channel are not charge injected but can still be readout for noise and pedestal measurement. Although the tested channel is not affected by directed injected charge, but there is a cross talk effect from the neighboring channel.

\begin{figure}
	\begin{center}
		\includegraphics[width=0.5\textwidth]{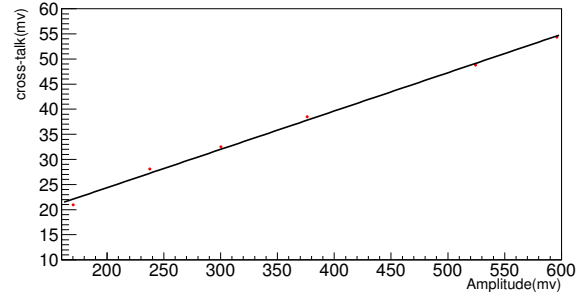}
		\caption{Relation of the strength of the cross-talk signal on neighbor channel and the injected charge.}
		\label{fig:crosstalk}
	\end{center}
\end{figure}

The cross-talk effect is first measured. Different charges are injected into one channel, and the PDO count of its neighboring channel is read out. A cross-talk signal proportional to the injected charge in neighboring channel is assumed. Comparing the charge injected and the readout from the neighboring channel, a straight line fit is applied as shown in Figure~\ref{fig:crosstalk}. The slope of 6.2$\%$ is taken as the cross talk efficiency.

The effect of cross-talk on non-neighboring channel can be ignored, which is verified with oscilloscope. 

\begin{figure}
	\begin{center}
		\includegraphics[width=0.52\textwidth]{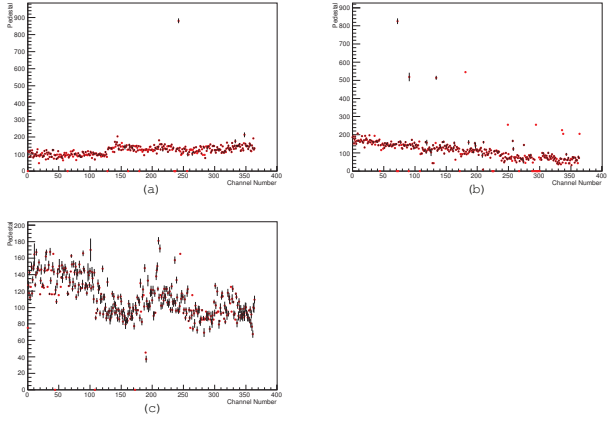}
		\caption{Pedestal and noise of sTGC detector with FEB connected(a).Pedestal and noise of first sTCG chamber(b).Pedestal and noise of second sTCG chamber(c).Pedestal and noise of third sTCG chamber.}
		\label{fig:noiseofeachlayer}
	\end{center}
\end{figure}

The PDO count from channels without charge injected are filled in histogram and fitted with Gauss function to obtain the pedestal(mean) and noise(sigma) of the channel. Considering the cross talk, the PDO count of the channel with injected charge times 6.2$\%$ are subtracted from the measured pedestal. The test result is shown in Figure~\ref{fig:noiseofeachlayer}, where the dots represent the pedestal of the channel, and the error bar represents the noise. The channels without error bar are dead channels. Some channels with very high pedestal are cross checked with the oscilloscope.

\subsection{Cluster of strip signals}

\begin{figure}
	\begin{center}
		\includegraphics[width=0.35\textwidth]{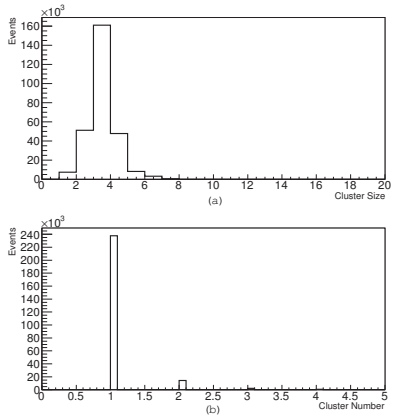}
		\caption{Distribution of ClusterSize(a) and ClusterNumber(b).}
		\label{ClusterDistribution}
	\end{center}
\end{figure}
When a muon hits a sTGC detector, the induced charge will be distributed on several adjacent strips, defined as a cluster. The center of gravity of a cluster is used to determine the hit position. 
In order to suppress noise, the clusters which contain only one channel are excluded. The number of clusters in one layer of the detector in one event is defined as ClusterNumber, the number of strips contained in a cluster is called ClusterSize as shown in Figure~\ref{ClusterDistribution}. 

\begin{figure}
	\begin{center}
		\includegraphics[width=0.52\textwidth]{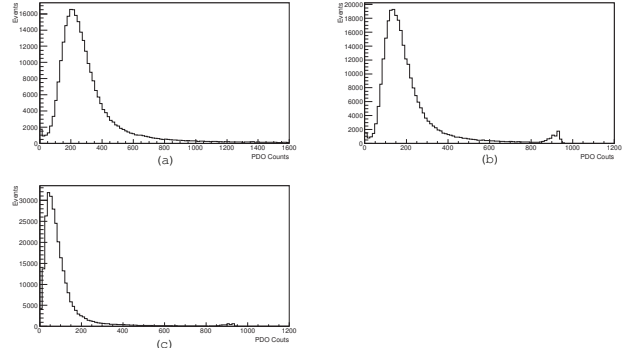}
		\caption{The charge of the peak channel in one cluster(a). The charge of the channel next to peak channel in magnitude in one cluster(b). The total charge of a cluster(c).}
		\label{fig:ChargeDistribution}
	\end{center}
\end{figure}

The digitized charge is represented by Q$_{i}$ where i is the channel number. Q$_{i}$ consists of three parts: the pedestal(P$_{i}$), noise($\sigma$$_{i}$) and signal charge. If Q$_{i}$$>$P$_{i}$+3*$\sigma$$_{i}$, the signal of that channel is defined as:

\begin{equation}
S_{i} = Q_{i} - P_{i}.
\label{eq:signalcharge}
\end{equation}
Otherwise, the signal of that channel is set to 0.

In one cluster, the peak signal is defined as the maximum signal of all the channels. The distribution of the peak signal is shown in Figure~\ref{fig:ChargeDistribution}(a). The distribution of the second peak signal in the cluster is shown in Figure~\ref{fig:ChargeDistribution}(b), and Figure~\ref{fig:ChargeDistribution}(c) shows the distribution of the amplitude of total signal of the cluster. The small peaks on the tails of the three plots are caused by electronic saturation. As shown in Figure~\ref{fig:ChargeDistribution}(c), the total charge in a cluster is about 250fC, which is 50$\%$ of the avalanche in one hit. This is consistent with the expected charge produced in an avalanche under the working voltage of 2900V\cite{ref:det2}.

\subsection{Position resolution}

To measure the position resolution, the hit position of one of the three chambers is compared to the expected position predicted by the other two chambers, assuming the muon travels along a straight path and the position resolution all over all the area of the chamber are uniform. Assuming that the hit position on three chambers are X$_{1}$, X$_{2}$, X$_{3}$ respectively, the expected position X$_{2C}$ predicted by X$_{1}$ and  X$_{3}$ is:
\begin{equation}
X_{2C} = \frac{L_{23}}{L_{12}+L_{23}}X_{1}+\frac{L_{12}}{L_{12}+L_{23}}X_{3},
\label{eq:PositionX2c}
\end{equation}
Where L$_{12}$ is the distance between the first and the second detectors and L$_{23}$ is the distance between the second and the third detectors.

\begin{figure}
	\begin{center}
		\includegraphics[width=0.45\textwidth]{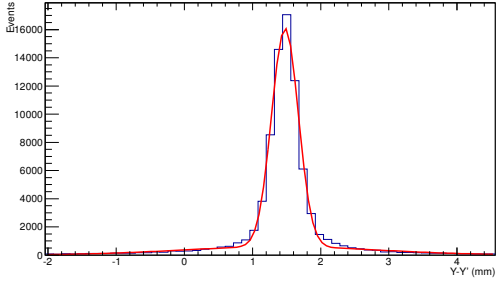}
		\caption{Difference between the actual position measured by strips of second chamber and the expected position predicted by the other two chambers.}
		\label{fig:MinusBefore}
	\end{center}
\end{figure}

The differences between X$_{2}$ and X$_{2C}$ fitted with a double Gaussian, representing the super-position of signal and background, are shown in Figure~\ref{fig:MinusBefore}. The fitted half width, $\omega$ , of the signal Gaussian function is 0.199mm. Using the error propagations:
\begin{equation}
\sigma_{2C} = (\sqrt{\frac{L_{23}^{2}}{(L_{12}+L_{23})^{2}}+\frac{L_{12}^{2}}{(L_{12}+L_{23})^{2}}})\sigma = k\sigma,
\label{eq:errorPositionX2c}
\end{equation}
\begin{equation}
\omega = \sqrt{\sigma^{2}+(k\sigma)^{2}} = \sqrt{1+k^{2}}\sigma,
\label{eq:width}
\end{equation}

and L$_{12}$ = L$_{23}$, so k is equal to $1/\sqrt{2}$, and the position resolution of the detector is:

\begin{equation}
\sigma = \frac{\omega}{\sqrt{1+k^{2}}} = \frac{\omega}{\sqrt{1.5}}=0.162mm.
\label{eq:positionresolution}
\end{equation}

The coordinates of the first channel of each layer are defined as 0, while the strips in the second layers are shifted by a half-width of the strip pitch (strip pitch 3.2mm) relative to the first and third layers, so the mean value is expected to be 1.6 mm. The mean of the signal Gaussian function is 1.48mm. The discrepancy between 1.48mm and 1.6mm comes from the uncertainty of manufacture of the strips.

\subsection{Relative rotation}

\begin{figure}
	\begin{center}
		\includegraphics[width=0.45\textwidth]{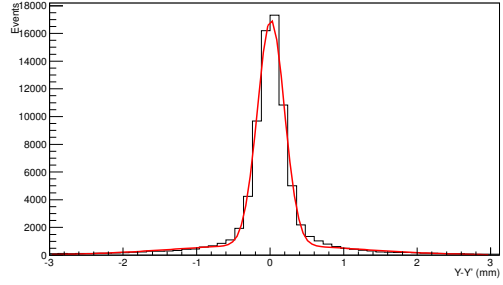}
		\caption{Different distribution of the actual position measured by strips of second chamber and the expected position predicted by the other two chambers after modified.}
		\label{fig:MinusAfter}
	\end{center}
\end{figure}
When the sTGC chamber is assembled, slight rotation and relative shift from the designed position can not be avoided. From the cosmic data, the relative rotation and shift is analyzed. As the relative rotation and shift of two parallel layers of chamber are not able to be concluded from cosmic muons of random directions, the relative rotation angle $\Psi$ around Z direction cross the centerl of the detector and shift Y$_{0}$ are defined for the second layer of chamber referring to the first and the third layers. When a muon passes through the three chambers, the hit position Y on the three chambers are measured by the sTGCs and X by the test bed. The difference between the X$^{'}$,Y$^{'}$ on the second chamber layer and the X,Y predicted by the first and the third chamber layers follow Gaussian distributions:

\begin{equation}
G(\Delta X) = \frac{1}{\sqrt{2\pi}\sigma_{x}}e^{-\frac{1}{2}(\frac{\Delta X}{\sigma_{x}})^{2}},
\end{equation}
\begin{equation}
G(\Delta Y) = \frac{1}{\sqrt{2\pi}\sigma_{y}}e^{-\frac{1}{2}(\frac{\Delta X}{\sigma_{y}})^{2}}.
\end{equation}

Where 
\begin{equation}
\Delta X = f(Y_{0},\Psi,X^{'},Y^{'})-X,
\label{eq:1}
\end{equation}
\begin{equation}
\Delta Y = f(Y_{0},\Psi,X^{'},Y^{'})-Y.
\label{eq:2}
\end{equation}

A likelihood function is composed,
\begin{equation}
\prod_{i}G(\Delta X_{i})G(\Delta Y_{i})  = L(Y_{0},\Psi).
\end{equation}

According to equation~\ref{eq:1} and ~\ref{eq:2}, we'll find that :
\begin{equation}
-{\log}L = \sum_{i}(\frac{\Delta X}{\sigma_{x}})^{2}+\sum_{i}(\frac{\Delta Y}{\sigma_{y}})^{2}.
\end{equation}

To minimize -$lgL$ by varying Y$_{0}$ and $\Psi$, the most likely Y$_{0}$ and $\Psi$ are obtained: $\Psi$ = 2.041e$^{-4}$ rad, Y$_{0}$ = 1.46mm, consistent with the mean of Gaussian function in Figure~\ref{fig:MinusBefore} .

Figure~\ref{fig:MinusAfter} reproduces the Figure~\ref{fig:MinusBefore} after the rotation and shift correction. The corrected position resolution is 0.152 mm. Mean of narrow Gaussian is 0.0113mm, verifying the analysis correction.

\subsection{Detection efficiency}

The detection efficiency of the sTGC can be measured by using the muon tracks reconstructed by chambers of the sTGC detectors. The number of muon tracks which are detected by all three chambers is defined as N$_{hit,3}$, and the number of muon tracks which are detected by at least the other two chambers is defined as N$_{hit,2}$. The detection efficiency is then defined as:
\begin{equation}
\Xi = \frac{N_{hit,3}}{N_{hit,2}}.
\end{equation}

\begin{figure}
	\begin{center}
		\includegraphics[width=0.4\textwidth]{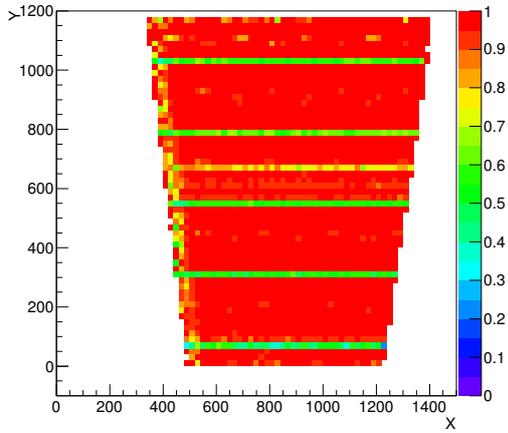}
		\caption{Typical detection efficiency of sTGC detector.}
		\label{fig:DetectionEfficiency}
	\end{center}
\end{figure}

For the purpose of scanning, the area of the second chamber is divided into grid of 2$\times$2 cm$^{2}$ squares and the efficiency in each square is calculated and shown in Figure~\ref{fig:DetectionEfficiency}. The dead area introduced by 5 supporting rulers in the chamber are visible clearly, at Y = 50mm, 300mm, 550mm,770mm, 1020mm respectively. Except for the low efficiency areas due to supporting rulers and dead electronic channels at Y = 660mm, the detection efficiency of the sTGC detector is uniformly above 95$\%$.

\section{Summary}

A prototype of the sTGC module of type S2 for the NSW project was successfully constructed in China, and the construction procedure is maturing with experience. The prototype was tested using cosmic rays and the performance was measured. A rotation angle of 2.041e$^{-4}$ rad of the detector validates the alignment method in the construction. The global position resolution 162$\mu$m without any correction and the detection efficiency of above 95$\%$ meets the requirement of the NSW.

\end{document}